\documentclass[12pt]{article}
\usepackage{amssymb}
\usepackage{amsmath,amsfonts,feynmp}
\usepackage{tipa}
\usepackage{stmaryrd}

\makeatletter\@addtoreset{equation}{section}\makeatother

\def\be{\begin{equation}}
\def\ee{\end{equation}}
\def\bea{\begin{eqnarray}}
\def\eea{\end{eqnarray}}

\makeatletter\@addtoreset{equation}{section}\makeatother

\renewcommand{\title}[1]{\vbox{\center\LARGE{#1}}\vspace{5mm}}
\renewcommand{\author}[1]{\vbox{\center#1}\vspace{5mm}}
\newcommand{\address}[1]{\vbox{\center\em#1}}

\begin{document}

\unitlength = .8mm

\begin{titlepage}
\begin{center}
\hfill \\
\hfill \\
\vskip 1cm

\title{ A Lifshitz Black Hole in Four Dimensional $R^2$ Gravity }

\vskip 0.5cm
 {Rong-Gen Cai\footnote{Email: cairg@itp.ac.cn}, Yan Liu\footnote{Email: liuyan@itp.ac.cn}} and {Ya-Wen
Sun\footnote{Email: sunyw@itp.ac.cn}}

\address{ Key Laboratory of Frontiers in Theoretical Physics
\\ Institute of Theoretical Physics, Chinese Academy of Sciences,
\\P.O. Box 2735, Beijing 100190, China}

\end{center}

\vskip 1.5cm

\abstract{We consider a higher derivative gravity theory in four
dimensions with a negative cosmological constant and show that
vacuum solutions of both Lifshitz type and Schr\"{o}dinger type with
arbitrary dynamical exponent $z$ exist in this system. Then we find
an analytic black hole solution  which asymptotes to  the vacuum
Lifshitz solution with $z=3/2$ at a specific value of the coupling
constant. We analyze the thermodynamic behavior of this black hole
and find that the black hole has zero entropy while non-zero
temperature, which is very similar to the case of BTZ black holes in
new massive gravity at a specific coupling. In addition, we find
that the three dimensional Lifshitz black hole recently found by E.
Ayon-Beato {\it et al.} has a negative entropy and mass when the
Newton constant is taken to be positive.}

\vfill

\end{titlepage}


\section{Introduction}
The anti-de Sitter/conformal field theory correspondence (AdS/CFT)
\cite{Maldacena:1997re} provides an elegant way to study strongly
coupled quantum field theories by relating them to certain classical
gravity theories. This holographic method is also quite useful in
describing condensed matter systems, for some nice reviews, see
\cite{Hartnoll:2009sz,
{Herzog:2009xv},{Faulkner:2009wj},{McGreevy:2009xe}} and references
therein. However, some condensed matter systems realized in
laboratories at their critical points are described by
non-relativistic conformal
 field theories (NRCFT). Thus it would be very useful to study the
 non-relativistic version of AdS/CFT to gain more knowledge of the strong
 coupling behavior of such condensed matter systems.

Non-relativistic conformal symmetry contains the scaling symmetry
\be\label{scaleinv} t\rightarrow \lambda^{z} t,
~~\textbf{x}\rightarrow \lambda\textbf{x} ,\ee where $z$ is the
so-called dynamical exponent. For $z=1$ this scaling symmetry comes
back to the familiar relativistic scale invariance. Such a
non-relativistic scale invariance (\ref{scaleinv}) can be exhibited
by either a Galilean-invariant theory or a Lifshitz-invariant
theory.
 In \cite{{Son:2008ye}, Balasubramanian:2008dm},
 the gravity dual for the Schr\"{o}dinger type field theory was proposed,
 while in \cite{Kachru:2008yh}, the gravity dual for the Lifshitz type field theory was proposed.

The thermal version of gauge/gravty dual is very useful in realistic
applications, so it is very interesting to heat up the previous
Schr\"{o}dinger vacuum and Lifshitz vacuum solutions proposed in
\cite{{Son:2008ye}, Balasubramanian:2008dm,{Kachru:2008yh}}. It is
easy to heat up the Schr\"{o}dinger vacuum and embed the thermal
solutions into string theory
\cite{{Herzog:2008wg},{Maldacena:2008wh},Adams:2008wt}, while,
however, it is very difficult to find black hole solutions which are
dual to the thermal Lifshitz field theory \cite{{Danielsson:2009gi},
{Mann:2009yx},Bertoldi:2009vn}. Some interesting attempts have been
made in \cite{{Brynjolfsson:2009ct},Balasubramanian:2009rx}. Other
interesting discussions on the gravity dual for  Lifshitz field
theory could be found in
\cite{{Adams:2008zk},{Taylor:2008tg},{Pal:2009yp},{Azeyanagi:2009pr},
{Ross:2009ar},{Bertoldi:2009dt},{Li:2009pf},Pang:2009ky,{Compere:2009qm}}.

Recently, in \cite{AyonBeato:2009nh} an interesting analytic black
hole solution which asymptotes to the Lifshitz vacuum solution was
found in the framework of the three dimensional New Massive Gravity
(NMG) \cite{Bergshoeff:2009hq} at a specific value of coupling. In
NMG, a special higher derivative gravitational term was added to the
usual action of general relativity and many interesting things may
happen at certain specific couplings
\cite{{Liu:2009bk},{Liu:2009kc},{Bergshoeff:2009aq},{Clement:2009gq},
{Nakasone:2009vt},{Nakasone:2009bn},{Liu:2009pha},{AyonBeato:2009yq},{Oda:2009ys},{Cai:2009ar},{Deser:2009hb},
{Kim:2009jm},{Clement:2009ka},{Oda:2009im},
{Oliva:2009ip},{Gullu:2009vy},{Chakhad:2009em},{Andringa:2009yc},Giribet:2009qz}.
This motivates us to look for higher dimensional asymptotically
Lifshitz black holes in pure higher derivative gravity system. At a
specific value of coupling, we indeed found an analytic
asymptotically Lifshitz black hole in four dimensions with the
dynamical exponent being $z=3/2$ and this solution could be viewed
as a generalization of the three dimensional case
\cite{AyonBeato:2009nh} to four dimensions. However, we did not find
any analytic Lifshitz black hole solutions for the higher derivative
gravity in five dimensions.

The black hole solution we found has very unusual thermodynamic
properties. Using the Wald entropy formula, we find that it has a
zero entropy while with a nonzero temperature which depends on the
mass parameter in the solution. This is very analogous to the
circumstance of the BTZ black hole in NMG at a specific
coupling~\cite{Liu:2009kc}. Though this does not necessarily mean
that the dual field theory is trivial, it is still mysterious in the
framework of pure gravity theory that a black hole has zero entropy
while with non-zero temperature.

In the next section, we give the analytic Lifshitz black hole
solution and the specific higher derivative gravity theory. In
Sec.~3, we discuss the thermodynamic properties of this black hole
and compare it with the BTZ black hole in NMG at a specific
coupling. In the last section, we give our conclusion and
discussions.

\section{The Lifshitz black hole solution}

Motivated by \cite{AyonBeato:2009nh} where an analytic
asymptotically Lifshitz black hole solution was found in NMG
\cite{Bergshoeff:2009hq}, which is a higher derivative gravity
system in three dimensions, we add higher derivative terms to the
four dimensional Einstein gravity theory and consider the following
action
\be\label{action}
 S=\frac{1}{16\pi G}\int d^4 x
\sqrt{-g}(R-2\lambda+\alpha R^2+\beta R_{\mu\nu}R^{\mu\nu}),
\ee
where $\alpha, \beta$ are coupling constants, $G$ is the four
dimensional Newton constant and $\lambda$ is the cosmological
constant. As we know in four dimensions the Gauss-Bonnet term is a
topological invariant and does not affect the equation of motion, so
this action (\ref{action}) is the most general form for gravity
theories with $R^2$ high derivative terms in four dimensions. The
corresponding equation of motion for the action (\ref{action}) is
\be
\label{eq}
G_{\mu\nu} +\lambda g_{\mu\nu}+Y_{\mu\nu}=0,\ee
where \bea G_{\mu\nu}&=&R_{\mu\nu}-\frac{1}{2}g_{\mu\nu}R,\nonumber\\
Y_{\mu\nu}&=&(2\alpha+\beta)(g_{\mu\nu}\nabla^2-\nabla_\mu\nabla_\nu)R+\beta\nabla^2G_{\mu\nu}\nonumber\\
   &&~~+2\alpha R(R_{\mu\nu}-\frac{1}{4}g_{\mu\nu}R)+2\beta(R_{\mu\rho\nu\sigma}-\frac{1}{4}
   g_{\mu\nu}R_{\rho\sigma})R^{\rho\sigma}.
 \eea
In the next subsections we will present Lifshitz and Schr\"odinger
vacuum solutions and Lifshitz black hole solutions for this action.

\subsection {Vacuum solutions}

We assume that the Lifshitz vacuum solution is of the form
\be\label{lifvac}
ds^2=-\frac{r^{2z}}{\ell^{2z}}dt^2+\frac{\ell^2}{r^2}dr^2+\frac{r^2}{\ell^2}(dx^2+dy^2),\ee
where $z$ is the dynamical exponent. We substitute this assumption
(\ref{lifvac}) into the equation of motion (\ref{eq}) and find that
the vacuum Lifshitz solution with the form (\ref{lifvac}) exists
only when the coupling constants and the cosmological constant
satisfy the following relation
 \bea \label{25} \beta\ell^{-2}&=&\frac{1-4\alpha\ell^{-2}(z^2+2z+3)}{2(2+z^2)} , \nonumber\\
\lambda\ell^2&=& -\frac{1}{2}(z^2+2z+3).\eea From the relation
(\ref{25}) we can easily see that to make sure that $z$ is a real
number, the cosmological constant should satisfy the condition
$\lambda\ell^2\leq-1$. In addition, we can have the ordinary AdS
vacuum solution with $z=1$ for $\lambda\ell^2=-3$ with arbitrary
value of $\alpha$ and $\beta$. \footnote{Note that the first
equation in (\ref{25}) is not valid for the case of $z=1$ as both
sides of that equation should be multiplied by a factor $z-1$. We
would like to thank E. Silverstein to point this out to us.}

The isometry group of this vacuum solution is generated by
\cite{{Kachru:2008yh},Adams:2008zk}
 \be M_{ij}=-i(x^i\partial_j-x^j\partial_i),~P_{i}=-i\partial_{i},~H=-i\partial_{t},
 ~D=-i(zt\partial_t+x^i\partial_i+r\partial_r)\ee
 which constitutes the Lifshitz symmetry algebra.
 The momentum $P_i$, Hamiltonian $H$ and angular momentum $M_{ij}$ enjoy
  the usual commutators, while the dilatation operator has nonvanishing
   commutators with the other generators as $[D,P_i]=iP_i$, $[D,H]=izH$, $[D,M_{ij}]=i(2-z)M_{ij}.$
It's natural to conjecture that the quantum gravity of (\ref{action})
 on the background (\ref{lifvac}) is dual to a $2+1$ dimensional
 non-relativistic quantum field theory which has a Lifshitz scale invariance with dynamical exponent $z$.

Interestingly the action (\ref{action}) also has a Schr\"{o}dinger
vacuum solution \be\label{schvac}
ds^2=-\frac{r^{2z}}{\ell^{2z}}dt^2+\frac{\ell^2}{r^2}dr^2+\frac{r^2}{\ell^2}(-2dtdx+dy^2),\ee
when \bea \beta\ell^{-2}&=&\frac{24\alpha\ell^{-2}-1}{2(2z^2-z-4)} , \nonumber\\
\lambda\ell^2&=& -3.\eea

Some analytic Schr\"{o}dinger black hole solutions are available in
the literatures
\cite{{Herzog:2008wg},{Maldacena:2008wh},Adams:2008wt}. We will
focus our attention on finding Lifshitz black hole solutions for the
action (\ref{action}) in the next subsection.

\subsection {The Lifshitz black hole solution}

Following \cite{AyonBeato:2009nh}, we assume that the metric of the
asymptotically Lifshitz black hole solution has the following form
\be
ds^2=-\frac{r^{2z}}{\ell^{2z}}F(r)dt^2+\frac{\ell^2}{r^2}H(r)dr^2+\frac{r^2}{\ell^2}(dx^2+dy^2),
\ee where $F(r)$ and $H(r)$ are functions depending on the radial
coordinate $r$ only. In order for the black hole solution to be
asymptotic to the Lifshitz vacuum solution (\ref{lifvac}), we demand
that these functions obey $\lim_{r\rightarrow \infty}
F(r)=\lim_{r\rightarrow \infty} H^{-1}(r)=1$. To make it a black
hole solution, we demand that $F(r)$ and $H(r)$  vanish at a given
radius $r=r_{H}$ where the horizon is located.

The equation of motion turns out to be solved by
\be
F(r)=H^{-1}(r)=1-\frac{r_{H}^3}{r^3},~z=\frac{3}{2}
 \ee
 when the
coupling constants and the cosmological constant take the following
value \be\label{cou}
 \alpha\ell^{-2}=1/33,\ \ \ \beta\ell^{-2}=0,\ \ \
\lambda\ell^2=-33/8.
\ee Here this solution is valid only at $z=3/2$
and at other values of the dynamical exponent $z$, we did not find
any analytic solutions.

Then the static asymptotically Lifshitz black hole for $z=3/2$ is
 given by
\be\label{lifbh}
ds^2=-\frac{r^3}{\ell^3}\big(1-\frac{r_H^3}{r^3}\big)dt^2
+\frac{\ell^2}{r^2}\frac{dr^2}{1-\frac{r_H^3}{r^3}}+\frac{r^2}{\ell^2}(dx^2+dy^2),\ee
and the corresponding gravity action is \be\label{ourac}
S=\frac{1}{16\pi G}\int d^4 x \sqrt{-g}(R-2\lambda+\alpha R^2),\ee
where $\alpha={\ell^2}/{33}$ and $\lambda=-33/8\ell^2$.

 It can be easily checked that the solution (\ref{lifbh}) has a curvature
singularity at $r=0$, where some scalar combinations of the Riemann
tensor diverge. The metric has a horizon at $r=r_H$, and the
boundary is localized at $r\rightarrow\infty.$ From the geometrical
point of view, it is indeed a black hole solution. In addition, this
black hole is asymptotically to the Lifshitz vacuum solution
(\ref{lifvac}) with $z=3/2$. It is natural to conjecture that the
quantum gravity of (\ref{action}) with the above special parameters
on the background (\ref{lifbh}) is dual to a $2+1$ dimensional
thermal non-relativistic Lifshitz quantum field theory with
dynamical exponent $z=3/2$. It is expected that this solution can
also describe the quantum critical region in condensed matter
systems at nonzero temperature.

Before discussing the thermal properties of this black hole, let us
 first have a look at the action (\ref{ourac}) at the
specific coupling we take here. At the specific coupling (\ref{cou})
the trace of the equation of motion (\ref{eq}) gives \be R=4\lambda
+6\alpha \nabla^2 R.\ee This is always satisfied by constant
curvature metrics with $R=4\lambda$. Then by substituting
$R=4\lambda$ back to the equation of motion (\ref{eq}) we find that
the equation of motion vanishes, which means that all constant
curvature solutions with $R=4\lambda$ are solutions of the equation
of motion (\ref{eq}) at the specific coupling (\ref{cou}). It is
easy to check that the black hole solution we found is just a
constant curvature solution of this kind.

\subsection{Unusual thermal properties}

Now let us focus on the thermal properties of the Lifshitz black
hole (\ref{lifbh}). The Hawking temperature of the black hole is
easy to find
\begin{equation}
T =\frac{3}{4\pi r_H}\left(\frac{r_H}{\ell}\right)^{5/2}.
\end{equation}
By Wald's formula for the black hole entropy~\cite{Wald},
\begin{equation}
S =- 2\pi \int_H \frac{\partial {\cal L}}{\partial
R_{abcd}}\epsilon_{ab}\epsilon_{cd},
\end{equation}
where ${\cal L}$ is the Lagrangian of a gravity theory, we find that
the Lifshitz black hole solution (\ref{lifbh}) has a vanishing
entropy in the action (\ref{ourac}). The vanishing entropy is
closely related to the fact that our solution satisfies the
relation: $1+2\alpha R=0$.

Indeed, all constant curvature solutions with $R=4\lambda$ are
solutions to the action (\ref{ourac}) and our solution is a constant
curvature solution with $R=-1/2\alpha =4 \lambda$. Furthermore, we
find that the action (\ref{ourac}) is always vanishing for our
solution, even when a boundary term is added
\begin{equation}
S_{\rm bt}=-\frac{1}{8\pi} \int_{\partial \cal M}d^3x
\sqrt{-h}(1+2\alpha R)K,
\end{equation}
where $K$ is the extrinsic curvature for the boundary hypersurface
${\partial \cal M}$ with induced metric $h$.  This implies that both
the free energy and mass of the black hole vanish as well
\begin{equation}
F=M=0,
\end{equation}
although the black hole has a nonvanishing horizon radius $r_H$.

Such thermodynamic behavior looks very strange. However, we find
that similar strange thermodynamic behavior also occurs for the
three dimensional Lifshitz black hole found in
\cite{AyonBeato:2009nh}. The action of NMG in three dimensions is
 \be \label{nmg}I=\frac{1}{16\pi
G}\int d^3x \sqrt{-g} \bigg[R-2\lambda_0-\frac{1}{m^2}K\bigg], \ee
where \be K=R^{\mu\nu}R_{\mu\nu}-\frac{3}{8}R^2,\ee $m$ is the mass
parameter of this massive gravity and $\lambda_0$ is a constant
which is different from the cosmological constant. The black hole
solution present in \cite{AyonBeato:2009nh} is
\begin{equation}
\label{3d} ds^2 = -\frac{r^{2z}}{\ell^{2z}}F(r) dt^2
+\frac{\ell^2}{r^2} H(r)dr^2 +\frac{r^2}{\ell^2}dx^2,
\end{equation}
where
\begin{equation}
F(r)=H^{-1}(r)= 1-\frac{r_H^2}{r^2},
\end{equation}
$z=3$, $m^2=-1/(2\ell^2)$, $\lambda_0=-13/(2\ell^2)$ and $r_H$ is an
integration constant. The black hole has a Hawking temperature
 $$T=\frac{1}{2\pi \ell}\left(\frac{r_H}{\ell}\right)^3,$$
 but a negative entropy
 $$ S= -\frac{A}{G},$$
 by the Wald's formula, where $A= r_H L/\ell$ and $L$ is the length
 of the coordinate $x$. By using the first law of black hole
 thermodynamics, $dM=TdS$, we find that the mass of the black hole
 (\ref{3d}) is negative
 \begin{equation}
 M = -\frac{L}{8\pi G\ell} \left(\frac{r_H}{\ell}\right)^4,
 \end{equation}
 if taking the solution (\ref{3d}) with $r_H=0$ as a vacuum solution
 being of vanishing mass.

 Here we should note that if we take the three dimensional
 Newton constant $G$ to be negative as in Topological massive gravity and NMG in
 asymptotically Minkowski spacetime, the entropy and the mass can be
 both positive. However, in usual gravity calculations concerned
 with black holes, we always take $G$ to be positive, so the
 thermal behavior of this three dimensional black hole is quite unusual
 compared with ordinary black holes.

\subsection{Remarks on the unusual thermodynamic behavior}

It seems quite unphysical that a black hole solution possesses a
zero entropy with arbitrary nonzero temperature, which is a property
of thermal Minkowski vacuum or thermal AdS spacetime, or at least
this may indicate that the corresponding field theory is trivial in
some sense. However, in fact, such properties for black hole have
also been found before in NMG for BTZ black holes, which does not
necessarily mean that the corresponding field theory is trivial
\cite{{Liu:2009bk},{Liu:2009kc},{Bergshoeff:2009aq}}. Let us have a
look at what happened there.

The action (\ref{nmg}) is the one for NMG in three dimensions and
the BTZ black hole is a solution to this action. The central charges
of NMG in the background of $AdS_3$ are \cite{Liu:2009kc}  \be
c_L=c_R=\frac{3\ell}{2G}\bigg(1-\frac{1}{2m^2\ell^2}\bigg).\ee The
entropy of the BTZ black hole is proportional to the central charges
and also has a factor $1-1/2m^2\ell^2.$ Thus at a specific coupling
$m^2\ell^2=1/2$, the entropy of the BTZ black hole vanishes while
the temperature of the black hole is nonzero. This is very analogous
to the condition we encounter here. In the case of NMG, it is shown
that at the specific coupling $m^2\ell^2=1/2$ under Brown-Henneaux
boundary conditions, the dual field theory is trivial. However, at
this specific coupling under a relaxed log boundary condition, it is
shown that the dual field theory is a log conformal field theory
which is not trivial though the BTZ black hole has a zero entropy.
On the gravity side this may indicate that there are other gravity
solutions besides the BTZ black hole which satisfy the boundary
conditions, such as pp wave solutions~\cite{AyonBeato:2009yq}. In
the case of this asymptotically Lifshitz black hole, it is expected
that there may be other gravity solutions which asymptotes the
Lifshitz vacuum solution under certain boundary conditions and the
whole system with all asymptotically Lifshitz solutions can describe
the dual condensed matter system at the quantum critical region.

\section{Conclusion and discussion}
In this paper we considered a higher derivative gravity system
(\ref{action}) in four dimensions with a negative cosmological
constant. We showed that vacuum solutions with both Lifshitz type
(\ref{lifvac}) and Schr\"{o}dinger type (\ref{schvac}) isometry with
arbitrary dynamical exponent $z$ exist in this action. We also
constructed an analytic black hole solution (\ref{lifbh}) which
asymptotes to the vacuum Lifshitz solution with $z=3/2$ at a
specific value of coupling. However, we found that this black hole
has an unusual thermodynamic behavior that it has zero entropy but
non-zero temperature.

This may seem unphysical or at least indicate that the dual field
theory may be trivial. However, we have an analogy of BTZ black
holes in NMG in three dimensions, where at a specific coupling, the
entropy of the BTZ black hole vanishes while the temperature is
non-zero. In that case, the dual field theory can be trivial or
nontrivial depending on the boundary conditions though the entropy
is always zero at that specific coupling. Thus in this
asymptotically Lifshitz case, we hope that
 there may also be a proper boundary condition under which the dual non-relativistic
 field theory is not trivial and can describe the condensed matter systems at the
  quantum critical region.  Of course this still needs further investigation to
   see whether there exist other asymptotically Lifshitz solutions or whether
   the dual field theory is trivial.

Despite that the dual field theory may not be trivial, it is still mysterious in
 the framework of a gravity theory that a black hole can have zero entropy while
  non-zero temperature.  Future work should be done to probe this black hole
  using various ways to get more information of this black hole or to find the
  origin of this unusual behavior.

  In addition, we found that the three dimensional Lifshitz black hole presented
  recently in \cite{AyonBeato:2009nh} has a positive Hawking
  temperature, but a negative entropy and a negative mass when
  we take the Newton constant to be positive. From the
  point of view of black hole thermodynamics, it is of great
  interest to further understand those unusual thermodynamic properties
  of black holes.

\section*{Acknowledgments}
This work was supported in part by a grant from the Chinese Academy
of Sciences with No. KJCX3-SYW-N2, grants from NSFC with No.
10821504 and No. 10525060.

\end{document}